\documentclass{elsart}
\usepackage[dvips]{graphicx}
\usepackage{amssymb} 

\def\MARU#1{\leavevmode \setbox0\hbox{$^\bigcirc$}%
\copy0\kern-\wd0 \hbox to\wd0{\hfil{#1}\hfil}}

\begin{document}

\begin{frontmatter}
\title{Directional scintillation detector for the detection of the wind of WIMPs}

\author[Physics]{\corauthref{cor1}Y.~Shimizu},
\ead{pikachu@icepp.s.u-tokyo.ac.jp}
\author[Physics]{M.~Minowa},
\author[Physics]{H.~Sekiya},
\author[ICEPP]{Y.~Inoue}

\address[Physics]{Department of Physics, School of Science, University 
of Tokyo, 7-3-1 Hongo, Bunkyo-ku, Tokyo 113-0033, Japan}

\address[ICEPP]{International Center for Elementary Particle Physics (ICEPP), University of Tokyo, 7-3-1 Hongo, Bunkyo-ku, Tokyo 113-0033, Japan}

\corauth[cor1]{Corresponding author.}


\begin{abstract}

The quenching factor for proton recoils in a stilbene scintillator
was measured with a $^{252}$Cf neutron source and was found to be 0.1 -- 0.17
in the recoil energy range between 300~keV and 3~MeV. 
It was confirmed that the light yield depends on the direction of 
the recoil proton.

The directional anisotropy of the quenching factor could be used 
to detect the wind of the WIMPs caused by 
the motion of the earth around the galactic center.

\end{abstract}

\end{frontmatter}

\clearpage

\section{Introduction}

There is substantial evidence that most of the matter in our Galaxy must be
dark matter and exist in the form of Weakly Interacting Massive 
Particles (WIMPs) \cite{balyon}. 
WIMPs can be directly detected through elastic scattering with nuclei in a
detector. The typical recoil energy is of order of 10~keV.
The detector is required to measure about 10 keV nuclear recoil energies and 
discriminate between nuclear recoils and $\gamma$-rays to reject background 
events.  

\begin{figure}[thb]
\includegraphics{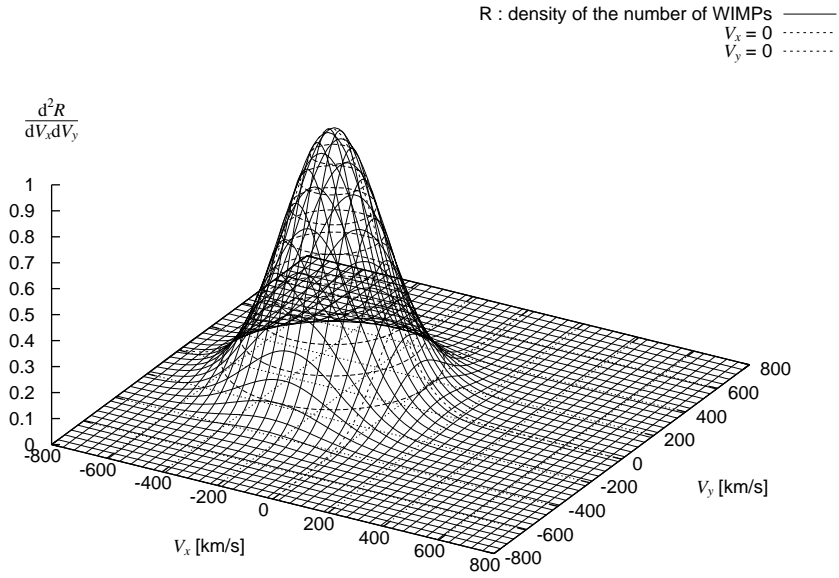}
\caption{Expected 2D velocity distribution of WIMPs relative to the earth. In this representation, the earth is moving toward the positive x direction with a velocity of 244 km/s. $V_{\rm x}$ and $V_{\rm y}$ are x and y components of WIMPs velocity $V$, respectively. RMS velocity of WIMPs is assumed to be 230 km/s.}
\label{wimp2d}
\end{figure}

For reliable detection of WIMPs, conventional searches depend mainly on
observing the annual 
modulation of the counting rate arising from the orbital motion of the 
earth around the sun and the
rotation of the solar system itself around the galactic center. 
However, this approach is subject to a small modulation amplitude and a large
systematic error 
induced by long-term changes of the condition of the detector and seasonal
variations of environmental radiations. 
For an alternative method, it might be useful to observe the wind of the 
WIMPs caused by the motion of the earth around the galactic center
\cite{dR/dE,diurnal} (Fig. \ref{wimp2d}). The effect of the wind is large 
($\simeq$ 244~km/s) and
a small systematic error can be achieved by means of periodically rotating 
the detector. 
To detect the wind of the WIMPs, a directional detector is required.

It is known that organic crystal scintillators such as anthracene 
and stilbene
have directional anisotropies in their scintillation responses to 
heavy charged particles \cite{anthHC,zp2}.
Protons and carbon ions are produced internally by WIMPs scattering and 
the scintillation responses depend on the direction of their motion. 
These anisotropies can be used to obtain directional information of the WIMPs
\cite{nonisotropic}.

In using scintillation detectors to observe WIMPs, one needs to measure 
the scintillation efficiency or the quenching factor of the nuclear recoil. 
For organic crystal scintillators, the anisotropies of the scintillation 
responses are expressed in the directional variations of the quenching factor. 

Measurements of the quenching factor for nuclear recoils have usually 
been performed with several MeV mono-energetic neutrons produced by nuclear 
reactions using accelerators \cite{qF-NaI,qF-CaF2}. 
The nuclear recoil was caused by elastic scattering with neutrons 
instead of WIMPs.
We performed the measurement 
employing a $^{252}$Cf neutron source which is easier to handle. 

In this paper, we report on the measurement of the quenching factor of a 
stilbene scintillator as a function of the recoil energy and its dependence
on the direction of the proton recoil. 

\section{Experiments}

The size of the stilbene crystal was 2 cm $\times$ 2 cm $\times$ 2 cm. 
It was purchased from Amcrys-H\footnote{URL: http://www.amcrys-h.com}.
Stilbene is an aromatic hydrocarbon including two benzene nuclei and 
forms a molecular crystal at room temperature which has an anisotropic 
structure.  

The crystal was covered with GORE-TEX\MARU{$^{\tt R}$}
to reflect the scintillation light and its cleavage plane was glued to
a PMT (H1161-50, Hamamatsu Photonics) with optical grease.

\begin{figure}[thb]
\begin{center}
\input{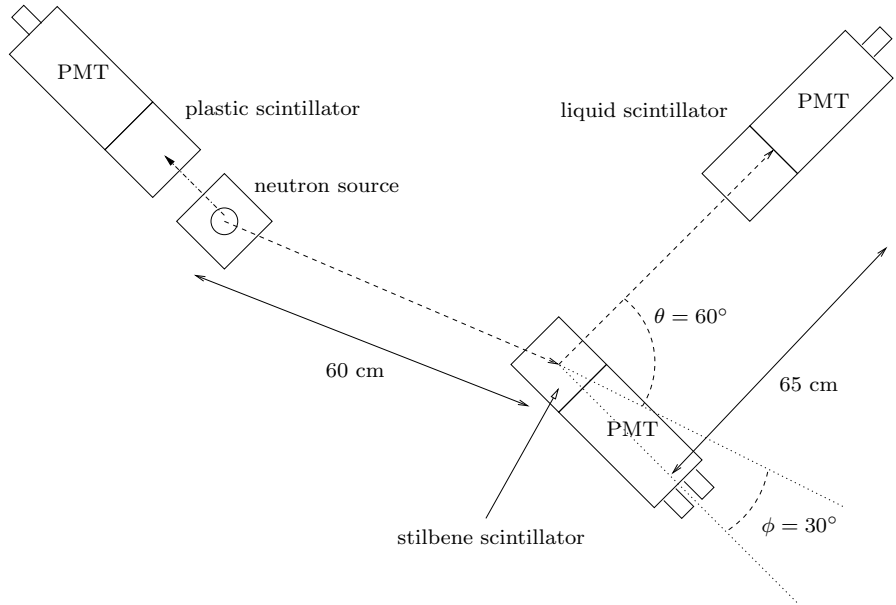}
\caption{Experimental setup.}
\label{arrangement}
\end{center}
\end{figure}

The experimental setup is shown in Fig. \ref{arrangement}.
We employed a 3.56~MBq $^{252}$Cf as a neutron source located 60 cm away 
from the stilbene scintillator.
  
The recoil energy $E_{\rm R}$ is calculated by the kinematics as

\begin{equation}
E_{\rm R} = \frac{2\left(1 + A - \cos^2 \theta - \cos
 \theta\sqrt{A^2 -1 +\cos^2 \theta}\right)} {(1 + A)^2}E_{\rm n},
\label{Er2}
\end{equation}

where $\theta$ is the scattering angle, $E_{\rm n}$ is the incident energy 
of the neutron and $A$ = (mass of nucleus)/(mass of neutron).

The scattering angle was determined by coincidence measurements between
the stilbene scintillator and a liquid scintillator located 65 cm away 
from the stilbene scintillator at 60 degrees. 
Since the neutrons from $^{252}$Cf source are not monochromatic but
have a continuous Maxwellian energy distribution,  
the incident energy of the neutron needs to be measured by the TOF method
between the source and the stilbene scintillator. 
To determine the timing of the nuclear fission of $^{252}$Cf, a prompt
$\gamma$-ray produced by the fission 
was detected with the plastic scintillator located by the neutron source.
The energy of the scattered neutron was 
measured by the TOF between the stilbene scintillator and the 
liquid scintillator. 

The quenching factor $Q$ is calculated by

\begin{equation}
Q = E_{\rm visible} / E_{\rm R},
\end{equation}

where $E_{\rm visible}$ is the measured electron-equivalent energy of the
stilbene scintillator as calibrated by $\gamma$-rays.

The recoil angle $\phi$ is calculated by

\begin{equation}
{\cos}^2\phi = \frac{1+A-{\cos}^2 \theta - \cos \theta \sqrt{A^2 - 1 + 
{\cos}^2 \theta}}{2A}.
\end{equation}  

For proton recoil, $\cos\phi = \sin \theta$.
We measured the quenching factor for two directions of the proton recoil, 
perpendicular and parallel to the cleavage plane.

\begin{figure}[thb]
\includegraphics{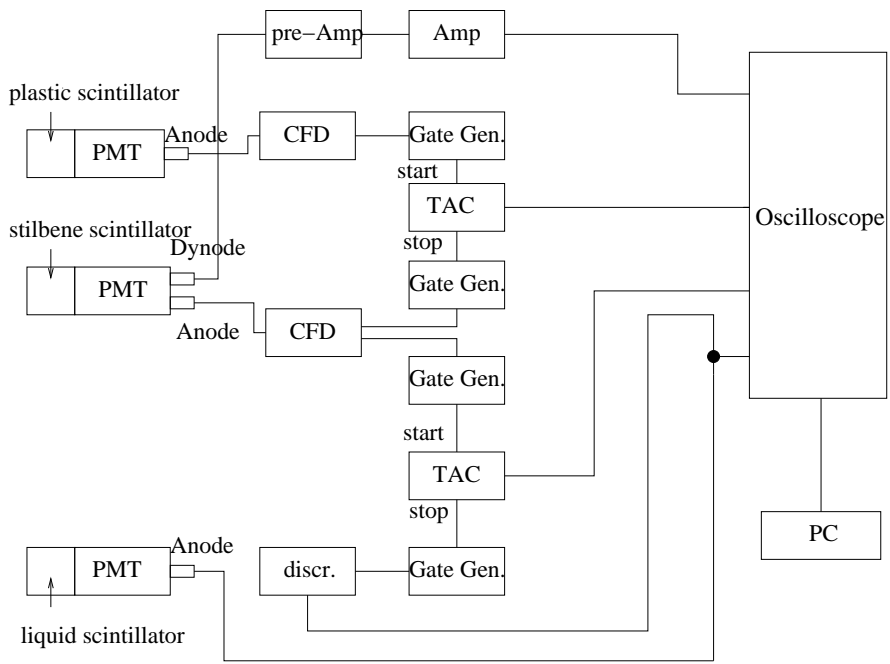}
\caption{Block diagram of the electronics for this experiment. Signals from an amplifier and two TACs were sensed and digitized by the oscilloscope and were sent to the PC.}
\label{setup}
\end{figure}

A schematic block-diagram of electronics for this experiment is shown 
in Fig.~\ref{setup}.
The pulse height signal of the stilbene scintillator was taken from
the dynode output of the PMT, while the timing signal was taken from
the anode output.
The timing signals of the three scintillators were used for start- and
stop-signals for two TACs. 
The pulses of the amplifier and the TACs were simultaneously recorded with a 
 4ch digital oscilloscope Tektronix TDS3034B.
The peak pulse heights of the signals were sensed and digitized by 
the oscilloscope and were sent to the PC using an RS232C serial line. 

\section{Analysis}

\begin{figure}[thb]
\includegraphics{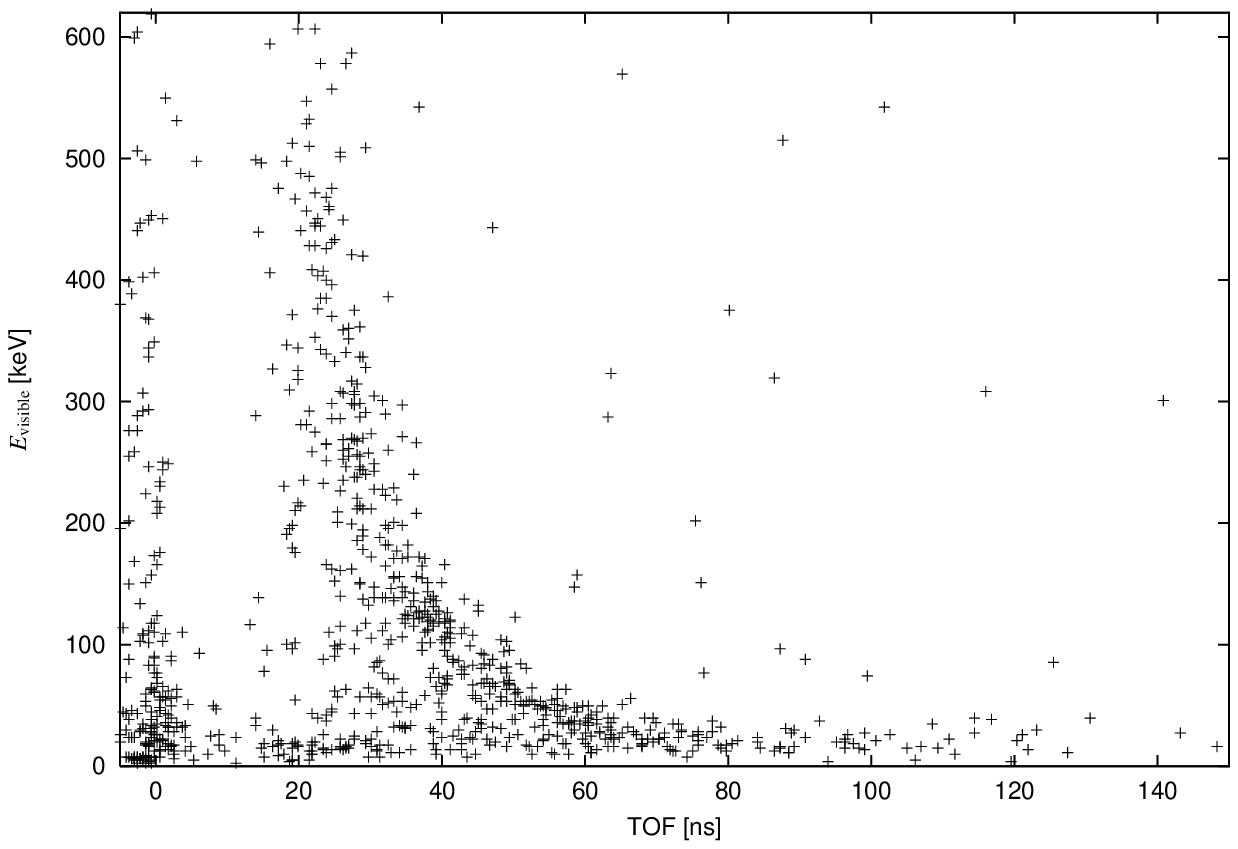}
\caption{Measured scatter plot of $E_{\rm visible}$ vs. TOF.}
\label{Ev-TOF}
\end{figure}

Fig. \ref{Ev-TOF} shows a scatter plot of the observed $E_{\rm visible}$ signal
versus TOF between the neutron source and the stilbene scintillator. 
The events around 0~ns are due to the $\gamma$-rays from the neutron source. 
The proton recoil events were distributed between 20 and 100~ns. 
Carbon recoil events have lower $E_{\rm visible}$, so that
they are not distinguished from background events.    

The recoil energy $E_{\rm R}$ can be calculated from eq. (\ref{Er2}) or by
subtracting the energy of the scattered neutron from the incident energy
calculated from the corresponding TOF. 
We used the latter method to obtain the recoil energy. 
The energy of the scattered neutron should be proportional to the incident 
energy as seen in
eq. (\ref{Er2}). We rejected background events which did not satisfy 
this relation within appropriate errors.  

\begin{figure}[thb]
\includegraphics{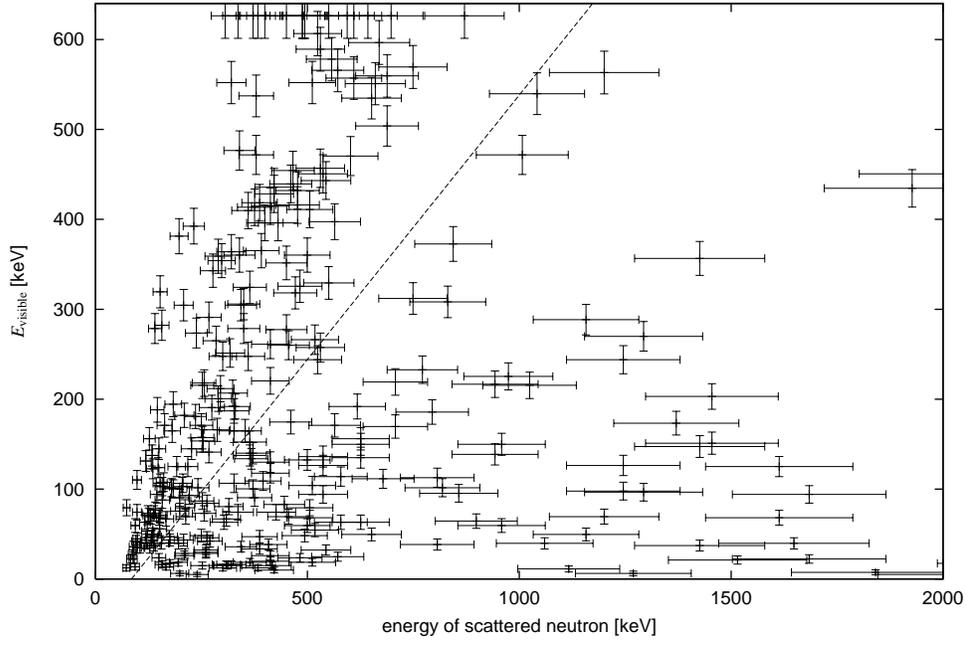}
\caption{Measured scatter plot of $E_{\rm visible}$ vs. energy of scattered neutron. We disregarded events below the dashed line which have large residual kinetic energies and small $E_{\rm visible}$ for proton recoils.}
\label{Ev-scattered}
\end{figure}

Fig. \ref{Ev-scattered} shows a scatter plot of the observed $E_{\rm visible}$ 
versus energy of the scattered neutrons. In case of the proton recoil, 
neutrons deposit 75\% of their energy in the stilbene, so that we disregarded
the events below the dashed line shown in Fig. 
\ref{Ev-scattered} which have large residual kinetic energies and small 
$E_{\rm visible}$ for the proton recoil.
These events were most probably induced by carbon recoils and accidental
coincidence background.

\begin{figure}[thb]
\includegraphics{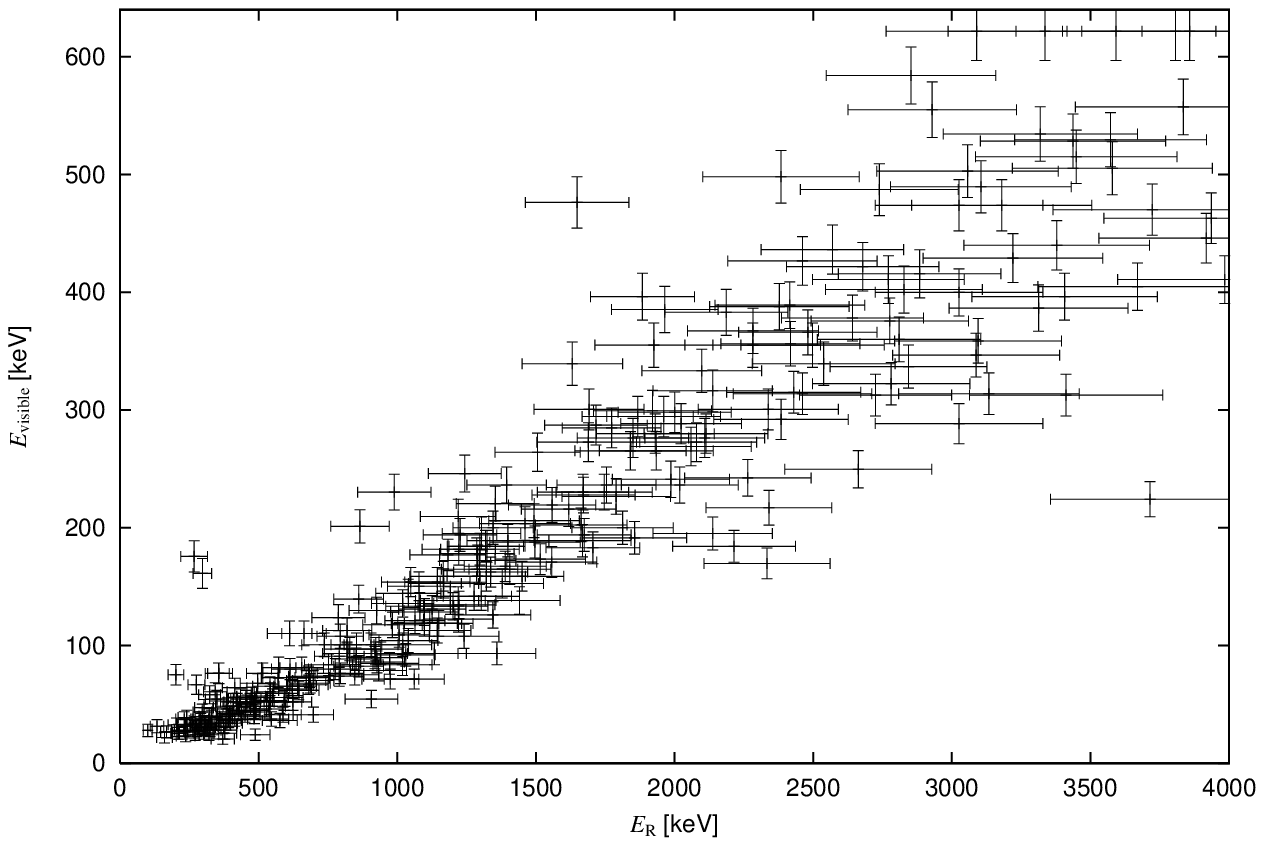}
\caption{Measured scatter plot of $E_{\rm visible}$ vs. $E_{\rm R}$.}
\label{Ev-Er}
\end{figure}

\begin{figure}[thb]
\includegraphics{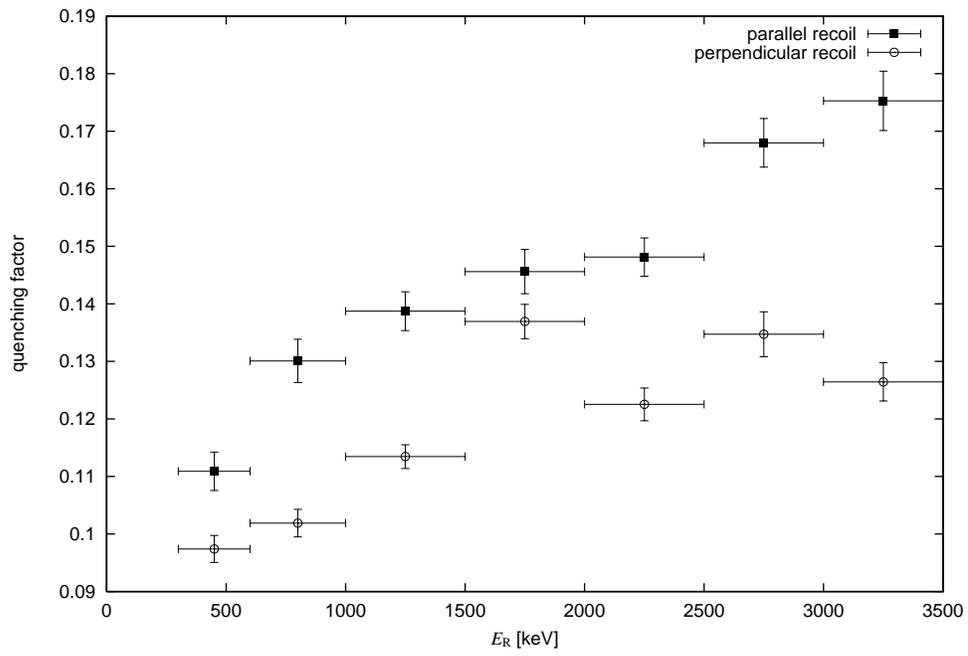}
\caption{Measured quenching factor for proton recoils on average as function of recoil energy. The quenching factor is smaller for the recoils perpendicular to the cleavage plane than for the recoils parallel to it.}
\label{QF}
\end{figure}

Fig. \ref{Ev-Er} shows a scatter plot of observed $E_{\rm visible}$ versus 
expected recoil energy $E_{\rm R}$.
Taking the average of $E_{\rm visible}/E_{\rm R} $ of the events 
in each bin of $E_{\rm R}$, the measured quenching
factor is shown in Fig. \ref{QF} as a function of the recoil energy. 
We measured the quenching factor for two recoil directions,  
parallel and perpendicular to the cleavage plane.
The former is shown by filled squares and the latter by open circles 
in Fig. \ref{QF}.

\section{Results and Discussions}

The quenching factor for proton recoils in a stilbene scintillator
was measured with a $^{252}$Cf neutron source and was found to range from 
0.10 to 0.17 with recoil energy from 300~keV to 3~MeV.
The measurement was performed for two recoil directions and we confirmed that
the quenching factor depends on the recoil direction. It was smaller for the
recoils perpendicular to the cleavage plane than for the recoils parallel 
to it. 
A measurement of the scintillation responses for $\alpha$-particles was 
reported in Ref. \cite{zp2}, where the responses for the two 
directions are qualitatively consistent with our result.

For the carbon recoil, definite data were not obtained because of the
low $E_{\rm visible}$ and the large background at low energies which 
were induced by accidental coincidences from $\gamma$-rays.
To measure the quenching factor in the low $E_{\rm visible}$ region,
we would need to improve the signal-to-background ratio.   

For the dark matter search, one can obtain directional information 
through comparing the data collected with different orientations of 
the detector.
We are planning to measure two energy spectra directing the crystal axes
of maximum and minimum scintillation responses toward the direction of the 
motion of the earth around the galactic center.

\clearpage

\end{document}